\documentclass{aa}  

\usepackage{xcolor}
\usepackage{natbib}
\usepackage{graphicx}
\usepackage{txfonts}
\usepackage[
  breaklinks = true,
   colorlinks = true,
  urlcolor   = blue,
  citecolor  = blue,
  linkcolor  = blue,
]{hyperref}

\newcommand{\tbold}[1]{{#1}}           

\begin{document} 

   \title{High-resolution observational analysis of flare ribbon fine structures}

    \author{
    Jonas Thoen Faber \inst{1,2}
    \and
    Reetika Joshi \inst{1,2}
    \and
    Luc Rouppe van der Voort \inst{1,2}
    \and
    Sven Wedemeyer \inst{1,2}
    \and
    Lyndsay Fletcher \inst{3,1}
    \and
    Guillaume Aulanier \inst{4,1}
    \and
    Daniel Nóbrega-Siverio \inst{5,6,1,2}
    }

   \institute{Rosseland Centre for Solar Physics, University of Oslo, P.O. Box 1029 Blindern, N-0315 Oslo, Norway\\
              \email{j.t.faber@astro.uio.no}
              \and
              Institute of Theoretical Astrophysics, University of Oslo, P.O. Box 1029 Blindern, N-0315 Oslo, Norway  
              \and
              SUPA School of Physics and Astronomy, University of Glasgow, Glasgow G12 8QQ, UK
              \and
              Sorbonne Universit\'e, Observatoire de Paris - PSL, \'Ecole Polytechnique, Institut Polytechnique de Paris, CNRS, Laboratoire de physique des plasmas (LPP), 4 place Jussieu, F-75005 Paris, France
              \and
              Instituto de Astrof\'isica de Canarias, E-38205 La Laguna, Tenerife, Spain
              \and
              Universidad de La Laguna, Dept. Astrof\'isica, E-38206 La  Laguna, Tenerife, Spain
             }

   \date{Received  / Accepted  }

 
  \abstract
   {
   Since the mechanism of energy release from solar flares is still not fully understood, the study of fine-scale features developing during flares becomes important for progressing towards a consistent picture of the essential physical mechanisms.
   }
   { 
   Solar flares release most of their luminous energy in the chromosphere. These luminous signatures, known as flare ribbons, act as the footpoints of the released energy and are crucial for the interpretation of reconnection processes causing these events. We aim to probe the fine structures in flare ribbons at the chromospheric level using high-resolution observations with imaging and spectral techniques. 
   }
   {
   We present a GOES C2.4 class solar flare (SOL2022-06-26T08:12) observed with the Swedish 1-m Solar Telescope (SST), the Interface Region Imaging Spectrograph (IRIS), and the Atmospheric Imaging Assembly (AIA). Utilising imaging data from SST, IRIS, and AIA, we detail both the global and fine-structure evolution of the flare. The high-resolution SST observations offer spectroscopic data in the H$\alpha$, \ion{Ca}{II}~8542~\AA, and H$\beta$ lines, which we use to analyse the flare ribbon.
   }
   {
   The flare was associated with a filament eruption. Fibrils and coronal loops were connected from a negative polarity region to two positive polarity regions. Within the eastern flare ribbon, chromospheric bright blobs were detected and analysed in \ion{Ca}{II}~8542~\AA, H$\alpha$, and H$\beta$ wavelengths. A comparison of blobs in H$\beta$ observations and \ion{Si}{IV}~1400~\AA\ has also been performed. These blobs are observed as almost circular structures having widths from 140 km--200 km. The intensity profiles of the blobs show a red wing asymmetry.
   }
   {
   From the high spatial and temporal resolution H$\beta$ observations, we conclude that the periodicity of the blobs in the flare ribbon, which are near-equally spaced in the range 330--550~km, is likely due to fragmented reconnection processes within a flare current sheet. This supports the theory of a direct link between fine-structure flare ribbons and current sheet tearing. We believe our observations represent the highest resolution evidence of fine-structure flare ribbons to date.
   }


   \maketitle


\section{Introduction}
\label{sec:introduction}

Solar flares are magnetically driven events that produce among the most energetic phenomena on the Sun \citep{1859MNRAS..20...13C, 1859MNRAS..20...15H} and have been routinely observed since the 1980s.
Flares are rapid events that occur commonly near an active region (AR) and release energies up 10$^{32}$\,erg producing emission throughout the electromagnetic spectrum leaving observational imprints across the solar atmosphere \citep{2008LRSP....5....1B, 2011SSRv..159...19F, 2011SSRv..158....5H}.
Though observed for a long time, the physical mechanism behind flares is still not fully understood.
The standard flare model explains the observed processes in the solar atmosphere as a response to magnetic reconnection \citep[see, e.g.,][]{2002A&ARv..10..313P, 2008LRSP....5....1B, 2011SSRv..158....5H}. 
The interpretation of this 2D model was initially formed based on a combination of theories (CSHKP model) by \cite{1964NASSP..50..451C, 1966Natur.211..695S, 1974SoPh...34..323H}, and \cite{1976SoPh...50...85K}.
The CSHKP model explains that an erupting flux rope creates an unstable region below it that becomes a site for reconnection.
This region is referred to as the current sheet where reconnection occurs, followed by the formation of parallel flare ribbons along the polarity inversion line (PIL). 
Later on, different numerical experiments explained the flare reconnection and ribbon formation in 3D \citep{2007ApJ...670.1453I, 2012A&A...543A.110A, 2013A&A...555A..77J, 2017A&A...601A..26Z, 2019MNRAS.490.3679W}.
Magnetic reconnection resulting in flares occurs typically in the corona \citep{1963ApJS....8..177P} where the magnetic field undergoes rapid reconfiguration.
Plasma in motion may provoke many sudden reconnection events within the current sheet, resulting in the conversion of magnetic potential energy into thermal energy and kinetic energy, including accelerated particles \citep{2002A&ARv..10..313P, 2019LRSP...16....3T}.
Particles propagate along the magnetic field and enter the dense chromosphere, where the majority of energy is deposited.
While the CSHKP model explains the processes of flare energy release on a macroscopic scale, it falls short of explaining the fine-scale structure and dynamics of the ribbons.
Observation resolving the reconnection processes in a flare current sheet is challenging as their signatures are typically produced in the X-ray domain \citep{2008LRSP....5....1B,2011SSRv..159...19F, 2020SoPh..295...75D}.
According to the CSHKP model, the chromosphere acts as a response to reconnection in the current sheet and is therefore often studied to infer the processes within the current sheet \citep[see, e.g.,][]{2002ApJ...565.1335Q, 2015ApJ...810....4B, 2024ApJ...965...16C}.
The flows around the current sheets lead them to be compressed into thin structures.
These are likely to be sites for fragmented and bursty reconnection \citep{2021ApJ...920..102W}.
Reconnection in the current sheet results in energy deposition in the chromosphere as energy flux travels along the magnetic fields and subsequently forms flare ribbons.
According to the CSHKP model, ribbons are formed in the chromosphere and act as footpoints where the majority of the energy is deposited.
They are typically recognised as elongated structures formed in pairs and associated with opposite polarities.
These flare ribbons are located at the periphery of high current density layers located at so-called quasi-separatrix layers \citep[QSL;][among many others]{1996A&A...308..643D, 1997A&A...325..305D, 2009ApJ...700..559M, 2016A&A...591A.141J, 2024A&A...687A.172J} that are narrow volumes in which magnetic tubes have a large squashing factor \citep[Q;][]{2002JGRA..107.1164T, 2007ApJ...660..863T}.
When observed at high spatial resolution, 
flare ribbons are seen to be composed of bright knots or plasma blobs moving along the ribbon \citep{2014ApJ...788L..18S, 2016ApJ...823...41D}.
These blobs are called ``kernels'' and formed due to energy deposition of nonthermal electrons \citep[see, e.g.,][]{1986lasf.conf...51K, 2021ApJ...920..102W, 2022ApJ...934...80L, 2023ApJ...954....7L, 2024A&A...690A.254G}.
\citet{2002ApJ...578L..91A} found that if the photospheric magnetic field is strong enough, kernels can be associated with hard X-ray sources.
Accelerated electrons are measured in the hard X-ray regime, which indicates that particles accelerate along magnetic fields to lower heights and consequently heat the ambient plasma.
The plasma enters the lower and denser atmosphere, where parts of the electron energy are transformed into radiation as particles collide.
The energy deposition in the chromosphere forming the ribbons can be seen as a response to magnetic reconnection in the corona \citep{2011SSRv..159...19F, 2021A&A...645A..80J}.
Flares are commonly observed in H$\alpha$ and \ion{Ca}{II}~8542~\AA\ \citep[see, e.g.,][and references therein]{2015ApJ...813..125K, 2020ApJ...896..120K, 2024MNRAS.527.5916M}.
\cite{2024A&A...685A.137P} utilised both spectral lines for observing an X9.3 class flare to \tbold{categorise} the different spectral properties that can be observed from flare ribbons.
\citet{2014ApJ...788L..18S} did a thorough analysis of the ribbon fine-structure of a C2.1 flare using observations from the New Solar Telescope of the Big Bear Solar Observatory.
They proposed that the fine-scale structures consisting of bright knots seen in H$\alpha$ were due to electrical resistivity being enhanced in the partially \tbold{ionised} plasma of the lower solar atmosphere.
\citet{2017NatCo...815905D} analysed electron beam simulations and found that during the onset of a flare, red-shifted H$\alpha$ profiles were formed similar to observations of a GOES C1.5 flare.
To our knowledge, fine-scale features in H$\beta$ are not commonly studied during flaring conditions, although observational studies of H$\beta$ ribbons have been done \citep[see, e.g.,][]{2017ApJ...850...36C, 2020ApJ...896..120K}.
The shorter wavelength of the H$\beta$ line compared to the H$\alpha$ line allows for observations at higher angular resolution in view of the change in diffraction limit for the same aperture.
The H$\beta$ line is, therefore, an interesting diagnostic tool that should be employed for studying the fine-scale structure of the flaring atmosphere. 
In this current study, we investigate the fine-scale structure and dynamics of flare ribbons utilising high-resolution observations from the Swedish 1-m Solar Telescope \citep[SST,][]{2003SPIE.4853..341S}.
The SST observations were coordinated with the Interface Region Imaging Spectrograph \citep[IRIS,][]{2014SoPh..289.2733D}, which expands the analysis to include diagnostics in the transition region (TR).
Full-disk observations by the Solar Dynamics Observatory \citep[SDO,][]{2012SoPh..275....3P} are also utilised, providing details in the corona and context of the observed AR.
In Sect.~\ref{sec:instruments_and_data_sets}, we explain the observing programs and present the datasets from SST, IRIS, and SDO.
Section~\ref{sec:analysis_and_results} describes the spectral data analysis and main observational findings.
We \tbold{finalise} the paper with a discussion in Sect.~\ref{sec:discussion} and a summary of our results in Sect.~\ref{sec:conclusions}.

\begin{figure*}[ht!]
   \sidecaption
   \includegraphics[width=12cm] 
    {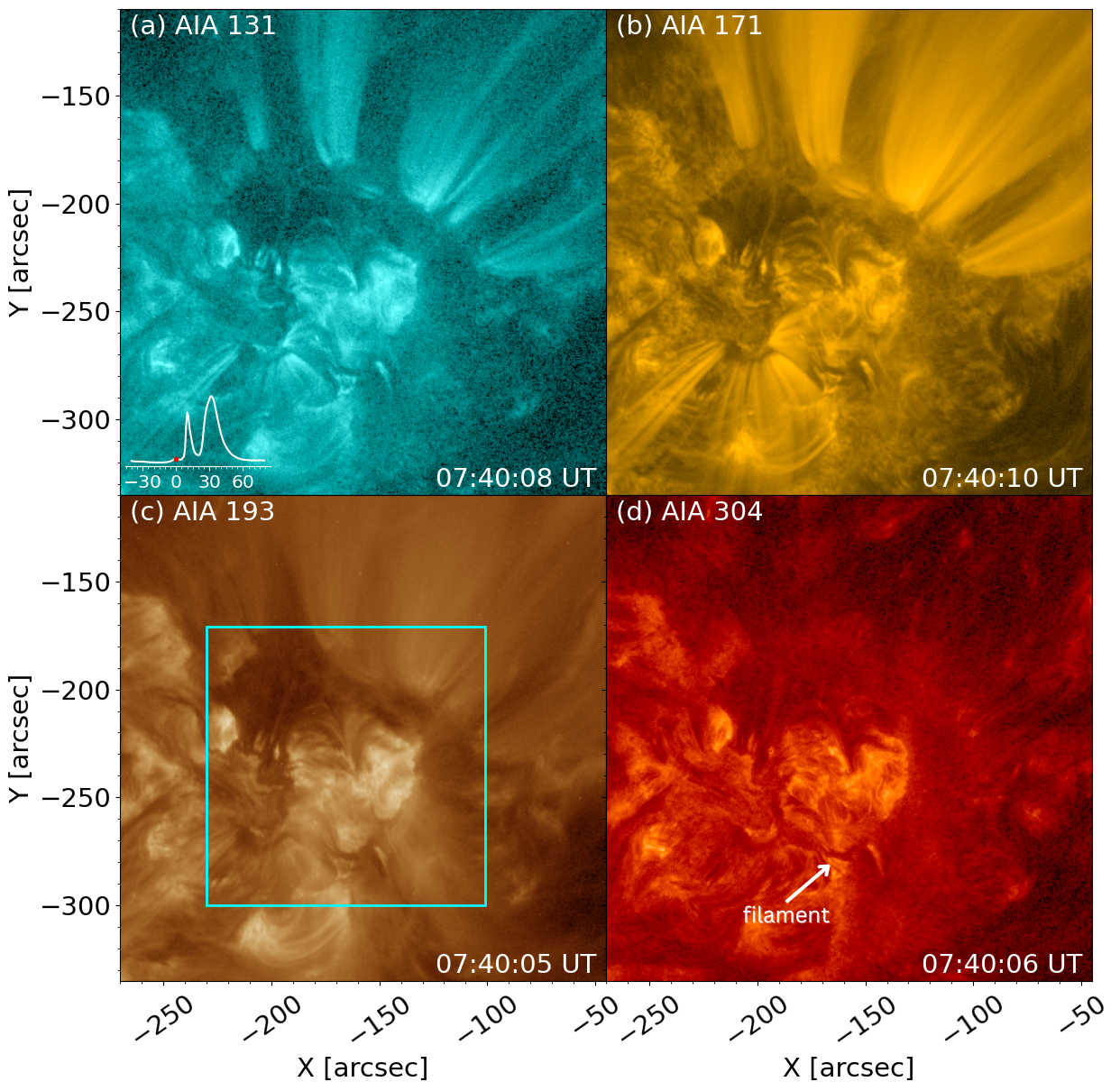}
    \centering
    \caption{
    Context images of NOAA AR 13040 on June 26, 2022, at 07:40 UT in different AIA channels.
    The GOES X-ray plot is superimposed in the lower left corner in panel~(a), where the red dot represents the time of the AIA images.
    The cyan rectangle in panel~(c) marks the FOV shown in Fig.~\ref{fig:sdo_context_pm200mf}.
    The white arrow in panel~(d) points to the erupting filament.
    This figure is associated with an animation that shows the full evolution of the filament eruption.
    }
    \label{fig:aia_context_filament_arrow}
\end{figure*}

\begin{figure*}[t!]
      \includegraphics[width=\textwidth]{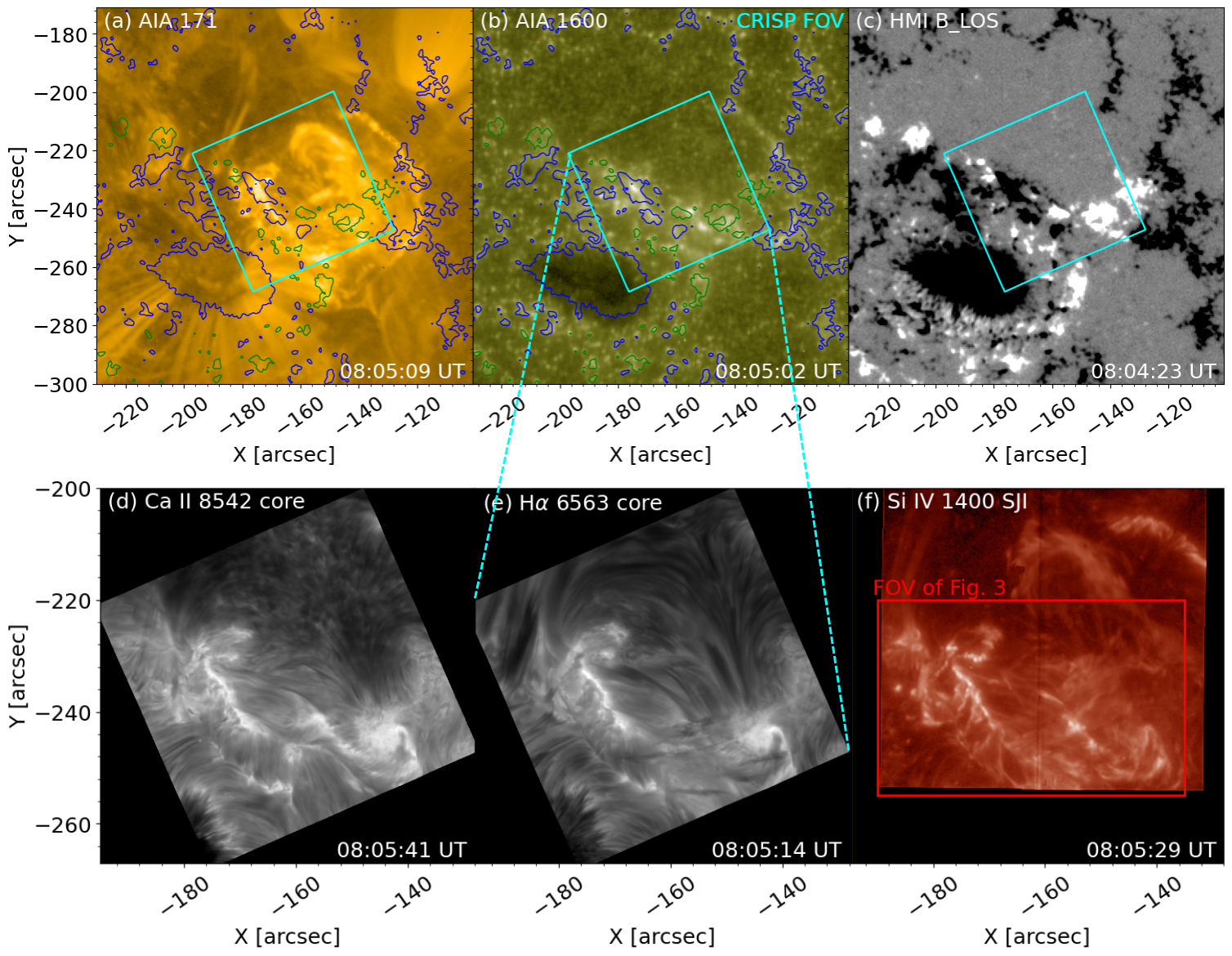}
    \centering
    \caption{
    The impulsive phase of the GOES C2.4 flare 5~min before the peak time.
    The larger FOV in the first row shows the global view of the flaring region in AIA and HMI observations.
    The pointing FOV of CRISP is overplotted in panels~(a)--(c) as cyan rectangles.
    The magnetic field from HMI is shown as contours at $\pm 200$~G in panels~(a) and (b), where green (blue) indicates positive (negative) polarity.
    Panel~(d) and (e) show the full FOV from CRISP in the \ion{Ca}{II}~8542~\AA\ line core and H$\alpha$ line core, respectively.
    Panel~(f) shows the aligned IRIS FOV in the \ion{Si}{IV}~1400~\AA\ channel.
    The red rectangle in panel~(f) shows the FOV as shown in Fig.~\ref{fig:sdo_sst_iris_aligned}.
    }
    \label{fig:sdo_context_pm200mf}
\end{figure*}


\section{Instruments and data sets}
\label{sec:instruments_and_data_sets}

\subsection{Overview of the flare observations}
\label{sec:flare_overview}

On June 26, 2022, two confined flares erupted in sequence outside the north-western edge of the penumbra in NOAA AR 13040, close to the disk centre.
They were classified as C1.9 (SOL2022-06-26T07:50) and C2.4 flares based on the GOES X-ray peak intensity within the 1--8~\AA.
The GOES start, peak, and end times of the C1.9 are 07:44~UT, 07:50~UT and 07:55~UT, respectively.
For the C2.4 flare, the times are 08:00~UT, 08:12~UT and 08:20~UT.
Images of the AR 10~minutes before the first flare provided by SDO are shown in Fig.~\ref{fig:aia_context_filament_arrow}. 
A plethora of coronal loops are evident in the AR, which constitute a complex system of magnetic fields.
The erupting filament shows as a dark elongated S-like structure in all coronal channels where one end is anchored at $[X,Y]=[-184\arcsec,-253\arcsec]$ in the penumbra and the other end at $[X,Y]=[-143\arcsec,-284\arcsec]$.
The first emission signatures become evident under the filament at the tip of the white arrow at 07:35:54~UT before the filament erupts at 07:43:06~UT.
The eruption results in a rising filament with a helical motion that produces the first flare within the AR.
As the filament stays in motion, a secondary bulk of strong emission becomes evident, and the second flare is identified. 
Two ribbons appear at $[X,Y]=[-175\arcsec,-235\arcsec]$ and $[X,Y]=[-145\arcsec,-245\arcsec]$ and are hereafter referred to as the eastern and western ribbon, respectively.
A movie showing the evolution of the AR and the production of both flares is available online.
SST and IRIS were pointing at AR 13040 ($[X,Y]=[-168\arcsec,-227\arcsec]$), northwest of the sunspot, effectively covering the eastern and western ribbons of the second flare.
\tbold{The field of view (FOV) of SST is outlined in Fig.~\ref{fig:sdo_context_pm200mf}a--c by a cyan rectangle}.
The majority of the first flare falls just south of both FOVs, whereas the second flare was well-captured during the onset, impulsive phase and gradual phase; hence, it is the focus of this paper.
%

\subsection{SDO}
\label{sec:sdo}

The Atmosphere Imaging Assembly \citep[AIA,][]{2012SoPh..275...17L} onboard SDO provides a time series of full-disk images covering the full duration of both flares. 
The extreme UV (EUV) channels (94~\AA, 131~\AA, 171~\AA, 193~\AA, 211~\AA, 304~\AA\ and 335~\AA) have a cadence of 12~s and are mostly sensitive to coronal temperature emissions, but also temperatures in the TR for the AIA 304~\AA\ channel.
The UV channels (1600~\AA\ and 1700~\AA) have a 24~s cadence and observe mainly the upper photosphere/lower chromosphere.
The 1600~\AA\ bandpass includes the \ion{C}{IV} line that is sensitive to TR temperatures as they occur during flares.
The pixel size of AIA is $0\farcs 6$.
The Helioseismic and Magnetic Imager \citep[HMI,][]{2012SoPh..275..207S} onboard SDO provides full-disk magnetograms derived from inversions of the \ion{Fe}{I}~6173~\AA\ line and the magnetic field of AR 13040 is shown in Fig.~\ref{fig:sdo_context_pm200mf}c.
The line-of-sight magnetic field by HMI has a cadence of 45~s and a pixel size of $0 \farcs 5$.
%

\subsection{IRIS}
\label{sec:iris}

A medium dense 16-step raster program was run with IRIS, which also provided slit-jaw images (SJI) in three different channels in far UV (FUV) and near UV (NUV).
The FUV SJI \ion{Si}{IV}~1400~\AA\ channel is dominated by the \ion{Si}{IV} lines, with the narrowband NUV 2796~\AA\ channel centred on the \ion{Mg}{II}~k core and the NUV 2832~\AA\ channel centred on the \ion{Mg}{II}~h wing.
The SJI images have a pixel scale of $0\farcs 167$, and the FOV is $65\arcsec \times 60\arcsec$.
The observing program pointed at $[X,Y]=[-166\arcsec,-225\arcsec]$ and lasted from 07:24~UT to 12:01~UT with a cadence of 21~s.
%


\subsection{SST}
\label{sec:sst}

The CRisp Imaging SpectroPolarimeter \citep[CRISP,][]{2008ApJ...689L..69S} is installed at the SST and was used for an observing program that includes the H$\alpha$, \ion{Ca}{II}~8542~\AA\ and \ion{Fe}{I}~6173~\AA\ spectral lines.
CRISP began observing at 07:36~UT and ended at 09:43~UT, covering both the impulsive phase and the gradual phase of the second flare.
The pixel size of CRISP is $0\farcs0586$, and the FOV coverage is approximately $53\arcsec~\times~52\arcsec$ with the pointing at $[X,Y]=[-153\arcsec,-235\arcsec]$.
The FOV of CRISP is shown as a \tbold{cyan} rectangle superimposed on the AIA and HMI images in Figs.~\ref{fig:sdo_context_pm200mf}a--c.
%
The H$\alpha$ program was carried out in spectral imaging mode and sampled 31 line positions between $\pm1500$~m\AA\ around the line core position with a wavelength separation of 100~m\AA. 
The \ion{Ca}{II}~8542~\AA\ line in spectropolarimetric mode was sampled at 20 line positions between -1680~m\AA\ and +2380~m\AA.
%
The sampling intervals were 80~m\AA\ around the core and increased from 120 to 700~m\AA\ in the wings.
The \ion{Fe}{I}~6173~\AA\ line was observed in spectropolarimetric mode to infer the magnetic field in the photosphere through Milne-Eddington inversions \citep{2019A&A...631A.153D}.
This line was sampled at 13 line positions between $\pm320$~m\AA\ with 40~m\AA\ steps around the core and 80~m\AA\ steps in the outer wings.
In addition, a continuum position was sampled at +680~m\AA.
The total cadence of the 3-line program with CRISP was 40~s.
%

The CHROMospheric Imaging Spectrometer \citep[CHROMIS,][]{2017psio.confE..85S} observes in the blue part of the spectrum.
It began observing at 08:04~UT, which is 28 minutes later than CRISP but 6 minutes before the GOES peak of the second flare.
The CHROMIS instrument was running a program that sampled the H$\beta$ line from $-$1300~m\AA\ to +1300~m\AA\ around the line core position with 100~m\AA\ increments.
Additional samplings were taken at $\pm2100$~m\AA, $\pm1950$~m\AA\ and +1500~m\AA\ covering a total of 27 wavelength positions resulting in a total cadence of 7~s.
The FOV is approximately $42\arcsec \times 67\arcsec$, and the pixel scale is $0\farcs 0379$.
Although CHROMIS began observing later than CRISP, omitting the pre-flare phase, the higher temporal and spatial resolution adds critical diagnostic information for the analysis of fine-scale components of the flare ribbons.
Degradation of data due to terrestrial atmospheric effects during the observing runs over longer periods is inevitable.
Some frames for our observations were affected by these effects during the lifetime of the flare.
The seeing quality of the data can be quantified in terms of the Fried parameter r$_0$, which was measured by the wavefront sensor of the adaptive optics system \citep{2019A&A...626A..55S, 2024A&A...685A..32S}.
The r$_0$ values varied between 4 and 35~cm for the near-ground seeing and between 4 and 14~cm for the full atmosphere. 
The CRISP and CHROMIS datasets were made science-ready through the SSTRED reduction pipeline \citep{2015A&A...573A..40D,2021A&A...653A..68L}. 
The SSTRED pipeline includes a seeing-correcting method utilising the Multi-Object Multi-Frame Blind Deconvolution \citep[MOMFBD,][]{2005SoPh..228..191V} image restoration procedure.
\tbold{
Most weight in the MOMFBD restoration of a spectral line scan is assigned to the images from the wideband channel that are acquired together with the narrowband images.
Both CRISP and CHROMIS have a wideband channel that branches off from the beam before the Fabry-P{\'e}rot interferometers and shows a scene that is dominated by the photosphere. 
This means that the impact of fast-evolving flare structures on the restoration process is minimal.
}


\subsection{Co-alignment of the observational datasets}
\label{sec:co-alignment}

This study is primarily focused on flare dynamics in the chromosphere, meaning SST observations are our principal data.
SST provides datasets with excellent spatial and temporal resolution, exposing fine-scale structures in the solar atmosphere.
Therefore, to retain the observational details from SST, both the SDO and IRIS data were re-scaled and interpolated onto the CRISP and CHROMIS pixel grid.
The SDO to SST alignment was done through a well-designed, open-source code written in Interactive Data Language (IDL) by Rob Rutten\footnote{The SDO to SST alignment procedure is available at \url{https://robrutten.nl/Recipes_IDL.html}.}.
The IRIS and SST co-alignment is based on cross-correlation between a subregion of the SJI \ion{Mg}{II}~h~2832 wing images and of the CRISP \ion{Fe}{I}~6173 wideband filter.
Both filters show a photospheric scene with granulation and AR structures.
The comparison comprises computing the X- and Y-offsets for each timestep in CRISP, where the SJI images were chosen based on a nearest-neighbour interpolation procedure.
The alignment between IRIS and CHROMIS undergoes the same procedure against the CHROMIS 4846~\AA\ wideband filter.
The wideband channel is not affected by H$\beta$ emission as the 6.5~\AA\ bandpass filter is centred 15~\AA\ away from the H$\beta$ line centre.
We estimate that the error of co-alignment is of the order of one IRIS pixel \citep[see][for more details]{2020A&A...641A.146R}.
The CRISP and CHROMIS narrowband images in the spectral line scans are aligned to their corresponding wideband channels through the MOMFBD restoration procedure.


\section{Analysis and results}
\label{sec:analysis_and_results}

\subsection{Morphology of the flare}
\label{sec:morphology_of_the_flare}

\begin{figure*}[t!]
    \includegraphics[width=\textwidth]{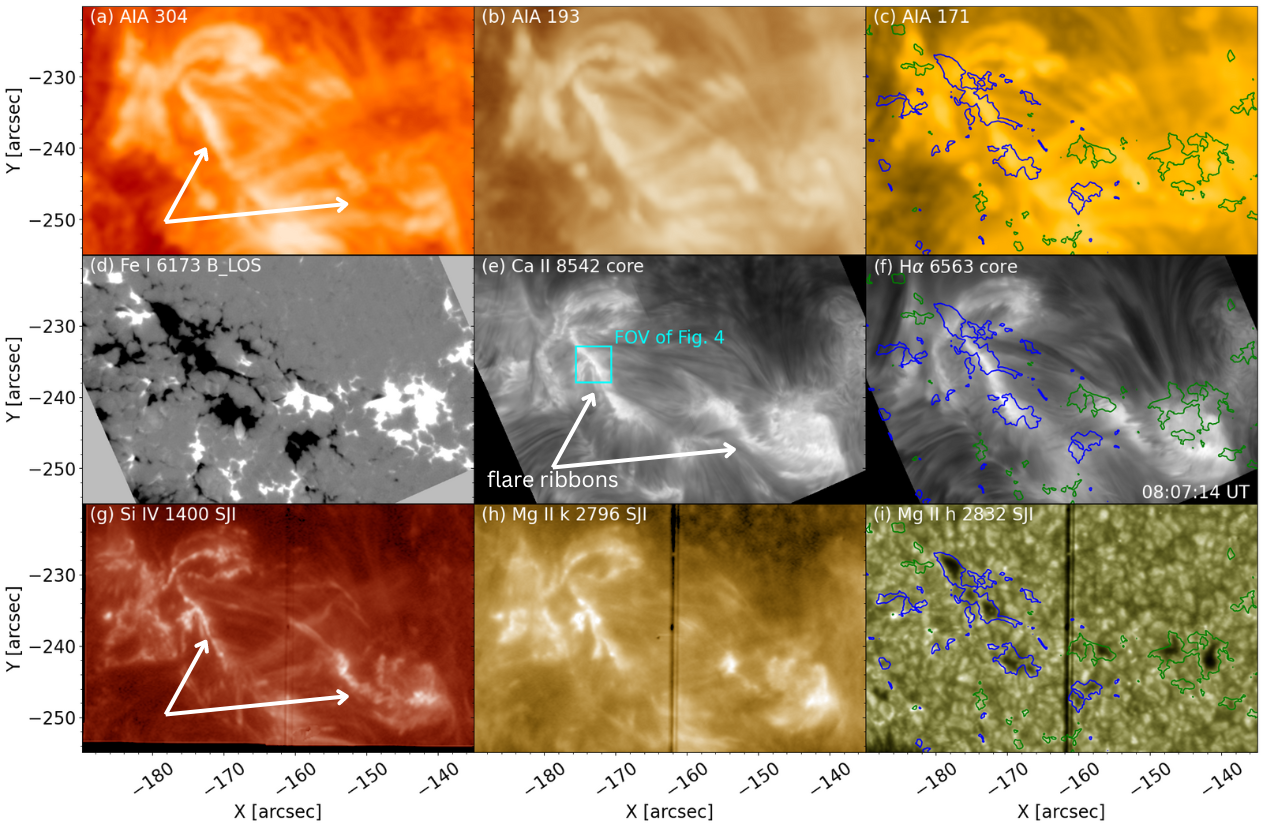}
    \centering
    \caption{
    Flare ribbon observation in different channels at 08:07~UT.
    Top row: (a) AIA 304~\AA, (b) AIA 193~\AA\ and (c) AIA 171~\AA\ channels.
    Second row: (d) line-of-sight magnetic field from \ion{Fe}{I}~6173~\AA\ line saturated at $\pm500$ G, (e) \ion{Ca}{II}~8542~\AA\ core and (f) H$\alpha$ core from CRISP.
    Bottom row: \ion{Si}{IV}~1400~\AA\ SJI, \ion{Mg}{II}~k~2796~\AA\ SJI and \ion{Mg}{II}~h~2832~\AA\ SJI from the IRIS telescope.
    White arrows in panels~(a), (e) and (g) point to the location of the flare ribbons.
    The magnetic field from the SST magnetogram is shown as contours in green and blue at $\pm 500$ G.
    The cyan rectangle in panel~(e) corresponds to the FOV, where fine-scale structures are studied.
    The dark vertical line in the middle of the IRIS images is the spectrograph slit.
    This figure is associated with an animation that is available online.
    }
    \label{fig:sdo_sst_iris_aligned}
\end{figure*}

AR 13040 was a source for many energetic events, such as jets, Ellermab bombs and flares.
The S-shaped filament erupted and triggered two recurrent C-class flares where the onset started at 07:44~UT.
During the first eruption, the filament was displaced in the direction further north but was still located in the outskirts of the sunspots before the second flare was triggered.
The pre-erupting filament is shown in the AIA~304~\AA\ channel in Fig.~\ref{fig:aia_context_filament_arrow}d, and a movie is available online.
The duration of the impulsive phase of the second flare was about 12~minutes and the gradual phase was about 18~minutes.
During the impulsive phase, two prominent ribbons emerged overlying a patch of strong and opposite-directed magnetic fields separated by an apparent north-south orientated PIL (see Fig.~\ref{fig:sdo_sst_iris_aligned} and associated animation).
The two ribbons are identified by the white arrows in Fig.~\ref{fig:sdo_sst_iris_aligned}e, where we differentiate them by the magnetic polarity in the photosphere.
Both polarities consist of pores in the photosphere and connecting fibrils in the chromosphere.
The eastern ribbon is overlying a patch of negative magnetic fields while fibrils on both sides of the straight part connect to patches of positive polarities.
The west side of the eastern ribbon shows fibrils connected to the western ribbon, while the east side is connected to a weaker positive field to the southeast.
This configuration suggests that the eastern ribbon is located at the end of a separatrix or a quasi-separatrix layer. 
The eastern ribbon was J-shaped, where the straight part stretched from
$[X,Y]=[-169\arcsec,-243\arcsec]$ to
$[X,Y]=[-176\arcsec,-231\arcsec]$
and from there morphed into a hook-shaped structure.
The hook extended perpendicular to the straight part as the flare developed.
The western ribbon can also be identified as a J-shaped structure, but the hook was more compact.
The structure of the ribbons was recognised in multiple EUV channels by AIA, implying that the ribbons reach coronal temperatures (see Fig.~\ref{fig:sdo_sst_iris_aligned}b--c).
Post-flare loops became evident during the gradual phase with the footpoints co-located with the ribbons.
Figure~\ref{fig:sdo_sst_iris_aligned} shows the atmospheric condition of the flare in the photosphere, chromosphere, TR and the corona during the impulsive phase.
We focus our study on the fine-scale structures in the chromosphere inside the \tbold{cyan} rectangle in Fig.~\ref{fig:sdo_sst_iris_aligned}e.
%


\subsection{Fine structures in flare ribbon}
\label{sec:ribbon_fine_structure}

\begin{figure*}[t!]
     \sidecaption
   \includegraphics[width=12cm]
    {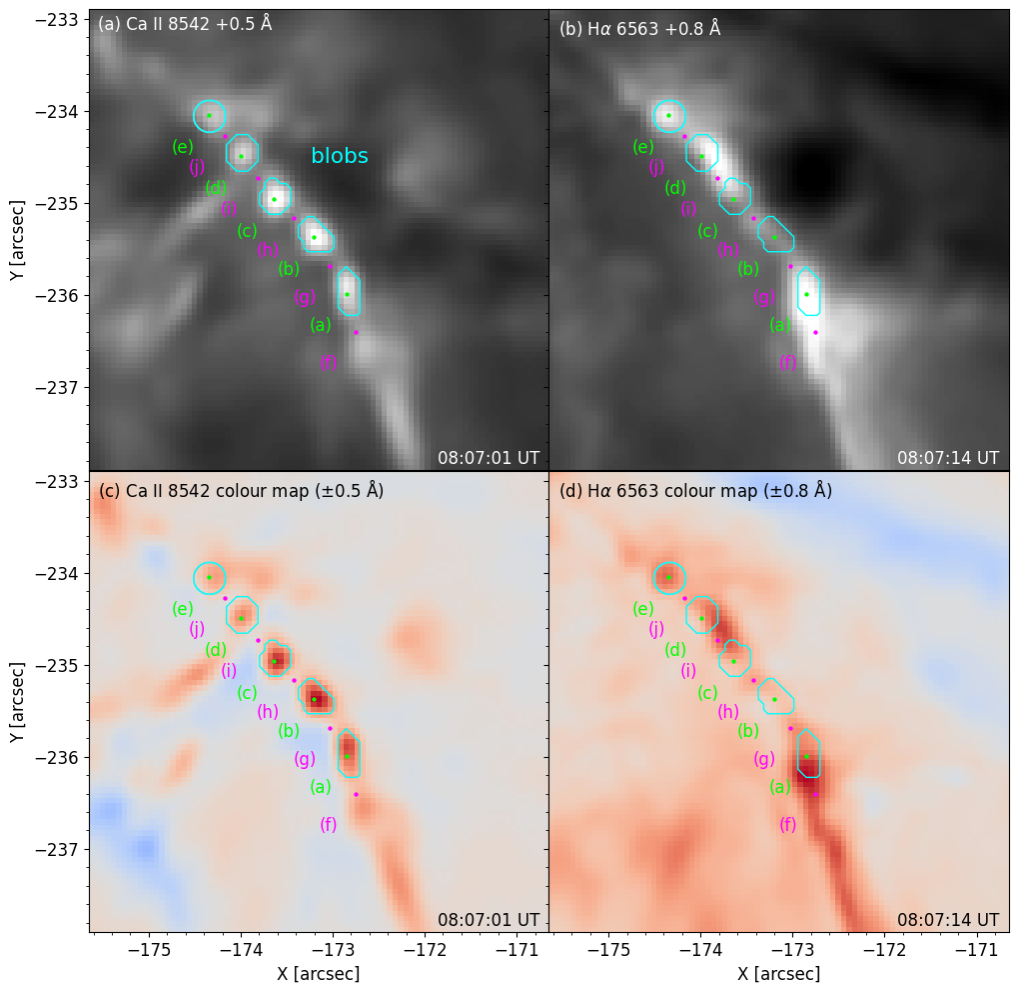}
    \centering
    \caption{
    Fine-scale features in the straight part of the eastern ribbon as outlined by the \tbold{cyan rectangle} in Fig.~\ref{fig:sdo_sst_iris_aligned}e.
    Upper row: CRISP images 
    in the \ion{Ca}{II}~8542~\AA\ and H$\alpha$ red wings.
    Bottom row: Colour maps computed by subtracting the wing intensities at the spectral positions shown in the titles.
    Both colour maps are saturated symmetrically around 0 so that pixels in white show no excess emission in either red or blue.
    Cyan contours in all panels outline bright features from panel~(a).
    Contours surrounding pixel~(a)--(d) were detected by the FWHM method, while the contour surrounding pixel~(e) was manually selected.
    The green pixels highlight the centre of the blobs and the magenta pixels are \tbold{located in} regions between the blobs.
    Pixel~(f) is selected below the lower-most blob in a region where brightening is evident for \ion{Ca}{II}~8542~\AA\ and H$\alpha$.
    All pixels are labelled with a letter a--j that corresponds to the panel titles in Fig.~\ref{fig:caii_blob_profile} and \ref{fig:ha_blob_profile}.}
    \label{fig:ha_caii_blobs_colourmap}
\end{figure*}

\begin{figure*}[ht!]
   \includegraphics[width=\textwidth]{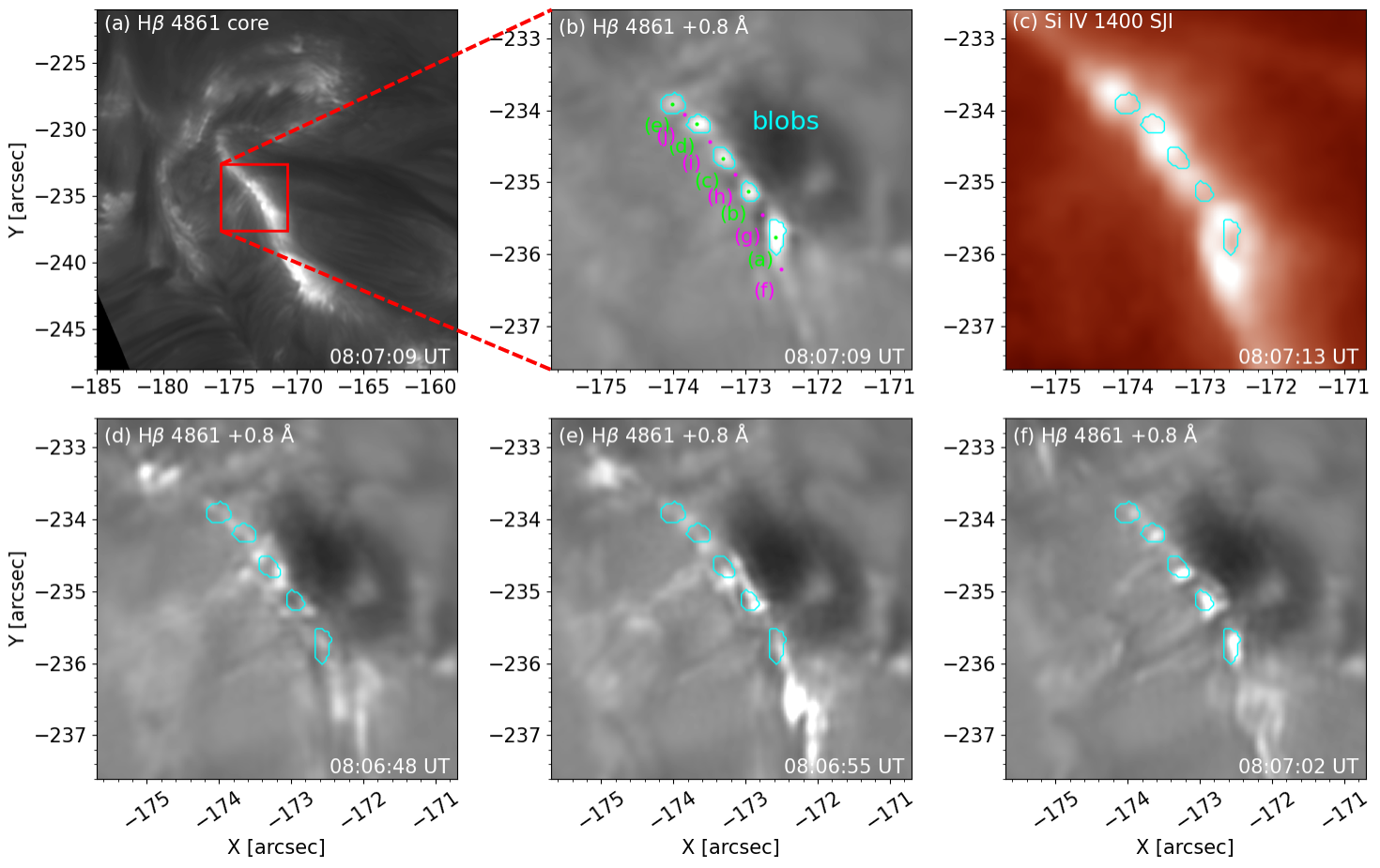}
    \caption{
    The eastern ribbon during the impulsive phase.
    Panel~(a) shows the full ribbon in H$\beta$ core and the red \tbold{rectangle} outlines the region that is shown at a larger magnification in panels~(b)--(f).
    Panel~(b) shows an image in H$\beta$~+0.8~\AA\ where the blobs are evident and panel~(c) shows the \ion{Si}{IV}~1400~SJI channel.
    Panels~(d)--(f) show H$\beta$~+0.8~\AA\ at different times before 08:07:09~UT.
    Cyan contours in panels~(b)--(f) outline the blobs from panel~(b) as detected by the FWHM method.
    The green and magenta dots mark the locations of the blobs and regions between the blobs that are used for spectral profile analysis.
    This figure is associated with an animation that is available online.
    }
    \label{fig:hbeta_siiv_blob_6panel}
\end{figure*}

CRISP provides high-resolution observations in the chromosphere, where most of the energy is deposited during a flare event.
In the straight part of the eastern ribbon, a line of five distinct and near equidistant bright features with varying peak intensities was detected in \ion{Ca}{II}~8542~+0.5~\AA, which are highlighted by the cyan contours in Fig.~\ref{fig:ha_caii_blobs_colourmap}a.
These features could be associated with flare kernels from the literature, but a clear definition is vague.
We refer to these structures as blobs based on the fine-scales from high-resolution observations, the spatial periodicity along the ribbon and their motion in a phase that is discussed in Sect.~\ref{sec:flare_blob_analysis}.
We emphasise that three of the features, meaning the blobs labelled (c)--(e), are close to circular structures while blobs~(a) and (b) are more elliptical.
The extent and perimeter of these blobs are defined according to the full-width at half-maximum (FWHM) of their spatial intensity profile. 
For that purpose, the maximum and minimum values are defined by the maximum intensity of a pixel within a local area and by the median intensity of a FOV, respectively.
We consider the median to be an effective way to represent a value that is not influenced by flaring pixels.
This method allows for a consistent definition of a blob, as the peak intensity will vary depending on the localised energy deposition.
We determine the location of each blob by the centre-of-mass method.
The blobs in Fig.\ref{fig:ha_caii_blobs_colourmap}a were measured to vary from 170--220~km (0\farcs 23--0\farcs 29) in widths and the separation between blobs varied from 430--520~km (0\farcs 59--0\farcs 71).
Although the algorithm was able to detect only four blobs in \ion{Ca}{II}~8542~+0.5~\AA, we identified a fifth faint blob located at $[X,Y]=[-174\farcs 4,-234\farcs 0]$ that is labelled (e). 
We argue that this feature should be included in the analysis as it is more prominent in other channels.
We then manually determined the location of this blob and highlighted the approximated width with a cyan circle.
Bright features are evident in the H$\alpha$~+0.8~\AA\ channel at the same region as the detected blobs in \ion{Ca}{II}~8542~+0.5~\AA\ but at a later time.
Blob~(e) that we manually identified in \ion{Ca}{II}~8542~+0.5~\AA\ is more prominent in this channel and shows a circular shape that is representative of a blob.
Additionally, two bright features in H$\alpha$ have elongated shapes that stretch parallel to the ribbon, so a total of three prominent features were detected.
The contours highlighting blob~(c) and (d) in Fig.~\ref{fig:ha_caii_blobs_colourmap}b are connected by one of the elongated features.
Two possible scenarios could be coalescing somewhere between the formation height of \ion{Ca}{II}~8542~\AA\ and H$\alpha$, or that the two \tbold{localised} bright points drift in time and consequentially merge.
We have identified \tbold{localised} bright points in \ion{Ca}{II}~8542~+0.5~\AA\ and H$\alpha$~+0.8~\AA\ along the straight part of the ribbon that does not share similar structures, except for the circular feature highlighted by (e).
We argue that this reveals fast-moving substructures in the low-chromosphere, which were unresolved between the scan of \ion{Ca}{II}~8542~+0.5~\AA\ and H$\alpha$~+0.8\AA.
Colour maps were made over a smaller FOV around the location of the blobs in \ion{Ca}{II}~8542~\AA\ (H$\alpha$) where the red wing intensity at +0.5~\AA\ (+0.8~\AA) is subtracted from the blue wing intensity at $-$0.5~\AA\ ($-$0.8~\AA), as shown in Fig.~\ref{fig:ha_caii_blobs_colourmap}c (Fig.~\ref{fig:ha_caii_blobs_colourmap}d).
Red (blue) colours indicate regions where the red (blue) wing component is stronger.
We call these maps colour maps rather than Doppler maps since the region contains both emission and absorption profiles.
A red-shifted emission profile will be red, while a red-shifted absorption profile will be blue.
Our main interest is the emission profiles in the ribbons. 
The colour maps show that both \ion{Ca}{II}~8542~\AA\ and H$\alpha$ profiles in the straight part of the ribbon are dominated by red wing enhancements during the impulsive phase.
Three of the blobs, labelled (a)--(c) in the \ion{Ca}{II}~8542~\AA\ colour map, show a significant red wing enhancement as compared to the rest of the ribbon, and is consistent with strong intensity levels in the red wing.
Similar behaviour can be identified in the H$\alpha$ colour map.
That is, areas along the ribbon dominant in the red wing are correlated with strong red wing intensities. 
%


\subsubsection{Plasma dynamics in H$\beta$}
\label{sec:chromosphere_dynamics_hbeta}

CRISP only observed the blobs in one frame, so to provide additional diagnostics of these events, CHROMIS will contribute to these events with a better temporal and spatial resolution.
The results are shown in Fig.~\ref{fig:hbeta_siiv_blob_6panel} and in the associated animation.
Five blobs were detected by the FWHM method in the H$\beta$ +0.8~\AA\ channel.
This includes an oval-shaped brightening at $[X,Y]=[-172\farcs 6,-235\farcs 6]$, as seen in Fig.~\ref{fig:hbeta_siiv_blob_6panel}b.
The width of the circular blobs varies from 140--200~km (0\farcs 19--0\farcs 27) and the variation of blob separation distances is 330--550~km (0\farcs 44--0\farcs 75).
The length of the semi-minor and semi-major axis of the oval-shaped blob is 140~km (0\farcs 19) and 340~km (0\farcs 46), respectively.
It was found that the location of the peak intensity of all blobs corresponds well with the centre-of-mass except for the oval-shaped blob, as the peak value is located $\sim 0 \farcs 18$ toward the north, which could be explained by projection effects.
The widths of the blobs extend over multiple CHROMIS pixels, as fixed at 0\farcs 0379.
Hence, we consider the blobs to be spatially resolved.
We aim to analyse the plasma dynamics that form the blobs in the H$\beta$ channel by investigating multiple frames.
The blobs in Fig.~\ref{fig:hbeta_siiv_blob_6panel}b are defined by the FWHM method and the contours are superimposed in Figs.~\ref{fig:hbeta_siiv_blob_6panel}b--f.
Five smaller fine-scale features surround the third and fourth contours, counting from the top left, as shown in Fig.~\ref{fig:hbeta_siiv_blob_6panel}d.
An exception is the feature underlying the third contour at $[X,Y]=[-173\farcs 3,-234\farcs 6]$, as it has an elongated structure along the y-axis.
The next scan exposes a drastic reconfiguration of individual fine-scale structures, as they can not be identified between the two scans.
The following scan in Fig.~\ref{fig:hbeta_siiv_blob_6panel}f also shows a significant change in the configuration of the bright features but exposes a plausible initiation of the blobs as the material begins to clump into distinct bright points that closely resemble the cyan contours.
The sequence of the three images before the blobs become evident shows that the bright features undergo a complex and rapid reconfiguration.
This dynamic evolution of the fine-scale structures suggests that a 7~s cadence does not sufficiently resolve the time evolution of the blobs.
%


\subsubsection{Blob analysis in H$\beta$ and \ion{Si}{IV}~1400}
\label{sec:flare_blob_analysis}

\begin{figure*}[t!]
    \includegraphics[width=\textwidth]{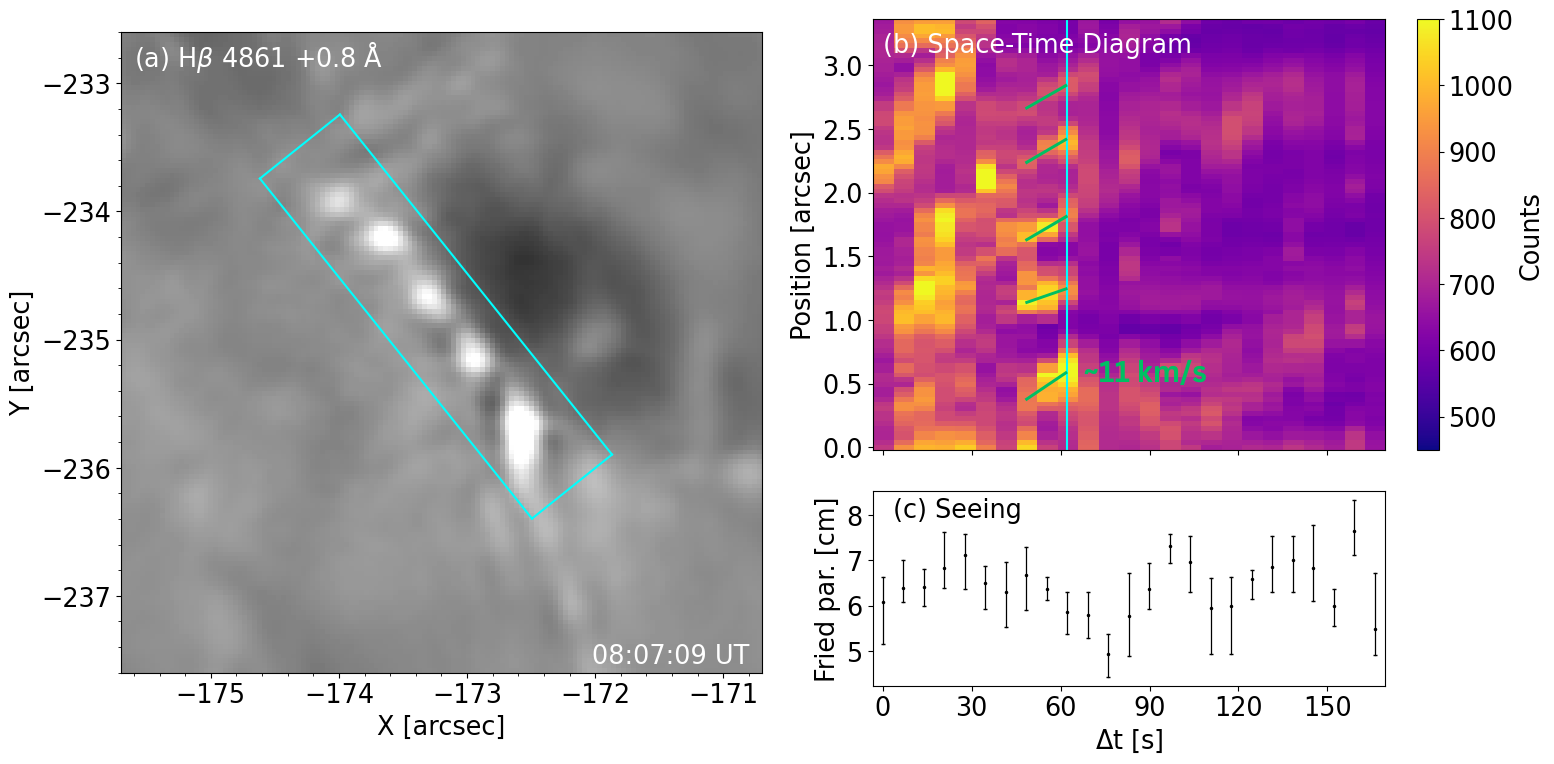}
    \centering
    \caption{
    Space-time diagram of the detected blobs from the H$\beta + 0.8$~\AA\ channel during the impulsive phase.
    Panel~(a) shows a FOV focused on the blobs.
    The FOV is presented as a red rectangle in Fig.~\ref{fig:hbeta_siiv_blob_6panel}a.
    The space-time diagram in panel~(b) is drawn from an artificial slit along the major axis of the cyan rectangle in panel~(a).
    The space-time diagram is determined based on the maximum value from the width of the artificial slit.
    The five green lines in panel~(b) mark the apparent slopes of the blobs.
    Panel~(c) shows the seeing quality as measured by the Fried parameter.
    The cyan line in panel~(b) highlights the same timestep as panel~(a)
    Panel~(b) and (c) share the x-axis in s from 08:05:59~UT.
    }
    \label{fig:hbeta_blobs_spacetime_diagram}
\end{figure*}

We created a space-time diagram from the H$\beta$ +0.8~\AA\ channel by using an artificial slit on top of the blobs as seen by the cyan rectangle in Fig.~\ref{fig:hbeta_blobs_spacetime_diagram}a.
The bright features move both parallel and perpendicular to the artificial slit.
To be able to track them as they move through this artificial slit, it is given an effective “width” by searching perpendicular to the slit.
Then the brightest feature within this range at each location defines, in effect, a new slit (it will not be a straight slit) and the intensity profile along this is used to construct the space-time diagram.
We found an optimal width of $0\farcs 8$ to sufficiently cover the fine-scale features, that is $\pm 0\farcs 4$ on each side of the artificial slit.
Note that the blobs may not be classified as blobs before 08:07:09~UT, but the diagram may give an idea of the source of any localised heating events.
The space-time diagram after this timestep shows that the fine-scale features become less prominent, but the weaker signals show trends of similar projections.
We investigate the blobs in Fig.~\ref{fig:hbeta_blobs_spacetime_diagram}b and the timestep of the detected blobs is marked by the vertical cyan line at 08:07:09~UT.
A vertical feature with varying intensities is evident, extending from $0\farcs 0$--$2\farcs 0$ and lasting from $\Delta$t = 7~s to $\Delta$t = 35~s.
This elongated feature is quickly fading at $\Delta$t = 35~s before the \tbold{initialisation} of multiple bright features becomes evident at $\Delta$t = 49~s.
Such \tbold{re-initialisation} of multiple bright points in a local area could indicate rapid reconnection that subsequently forms the blobs.
The bright features that are seemingly connected with the blobs show a plausible systematic motion along the straight part towards the hook-end of the ribbon.
While bright features are not identified as blobs in earlier time frames, these individual features seem to form slopes in the space-time diagram that extend over 2--3 timesteps.
Four near-parallel slopes are identified at approximate positions 0.5, 1.3, 1.7, and 2.3, with an additional faint slope at 2.8.
These slopes are drawn as green slanted lines in the space-time diagram for reference, as seen in Fig.~\ref{fig:hbeta_blobs_spacetime_diagram}b, and were used to estimate the apparent velocities.
We computed the average apparent velocity based on these slopes to be 11~km~s$^{-1}$.
%


%
The flare ribbon can be identified in \ion{Si}{IV}~1400~\AA\ as the bright structures correspond well with the ribbon seen in the H$\beta$ line core (see white arrows in Fig.~\ref{fig:sdo_sst_iris_aligned}g).
Although IRIS is not able to resolve the blobs at the same level as CHROMIS, equidistant bright features are identified in the \ion{Si}{IV}~1400~\AA\ images.
Three features were identified at 
$[X,Y]=[-174\farcs 2, -233\farcs 8]$, 
$[X,Y]=[-173\farcs 5, -234\farcs 4]$, 
$[X,Y]=[-172\farcs 6, -235\farcs 8]$ 
and a less prominent feature at 
$[X,Y]=[-173\farcs 1, -235\farcs 0]$ in Fig.~\ref{fig:hbeta_siiv_blob_6panel}c.
The extent of these locations corresponds to a region where stronger emission is evident along the straight part of the ribbon.
These bright features show tendencies of propagating towards the hook-end of the ribbon.

\tbold{
The IRIS slit did not cross the part of the ribbon where the blobs are located.
This makes it impossible to analyse these features in the IRIS spectral lines.
}


\subsection{Spectral analysis of fine structures}
\label{sec:blob_spectral_analysis}

\subsubsection{Spectral line profiles in \ion{Ca}{II} and H$\alpha$ observations}
\label{sec:crisp_profiles}

\begin{figure*}[t!]
    \includegraphics[width=\textwidth]{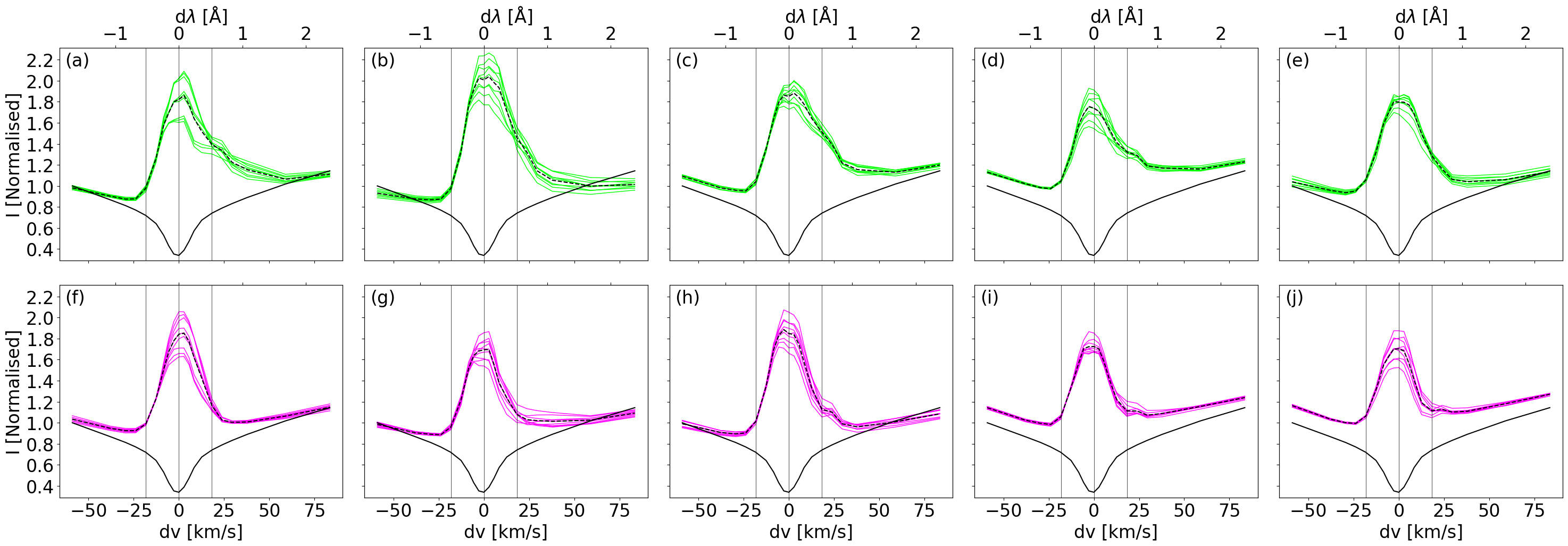}
    \centering
    \caption{
    Spectral profiles in the \ion{Ca}{II}~8542~\AA\ channel along the flare ribbon.
    The upper row \tbold{shows} profiles from blob locations in green and the bottom row \tbold{shows} profiles from regions between the blobs in magenta.
    \tbold{Positive (negative) d$\nu$ corresponds to red-shifts (blueshifts).}
    Each panel is colour-coded and given a title in accordance with the pixels in Fig.~\ref{fig:ha_caii_blobs_colourmap} and presents a $3 \times 3$-grid of adjacent pixels.
    In every panel, the solid coloured lines represent observed spectral profiles, the black dashed line represents the average, and the black solid line is the average profile over a QS region from the same timestep.
    All profiles are normalised by the blue wing of the QS average profile.
    The middle black vertical line highlights the core positions, while the left and right vertical lines show the wavelength positions used to create the Doppler map in Fig.~\ref{fig:ha_caii_blobs_colourmap}c.
    }
    \label{fig:caii_blob_profile}
\end{figure*}

\begin{figure*}[t!]
    \includegraphics[width=\textwidth]{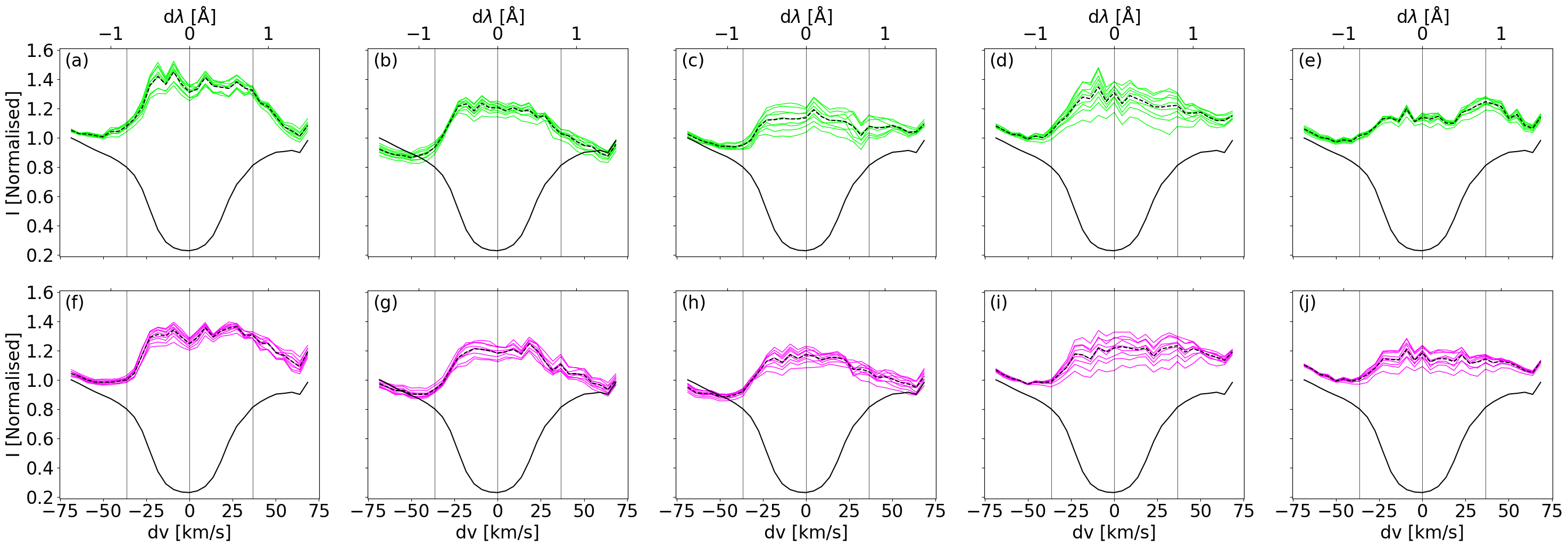}
    \centering
    \caption{
    Same as Fig.~\ref{fig:caii_blob_profile}, but for H$\alpha$.
    }
    \label{fig:ha_blob_profile}
\end{figure*}

\begin{figure*}[ht!]
    \includegraphics[width=\textwidth]{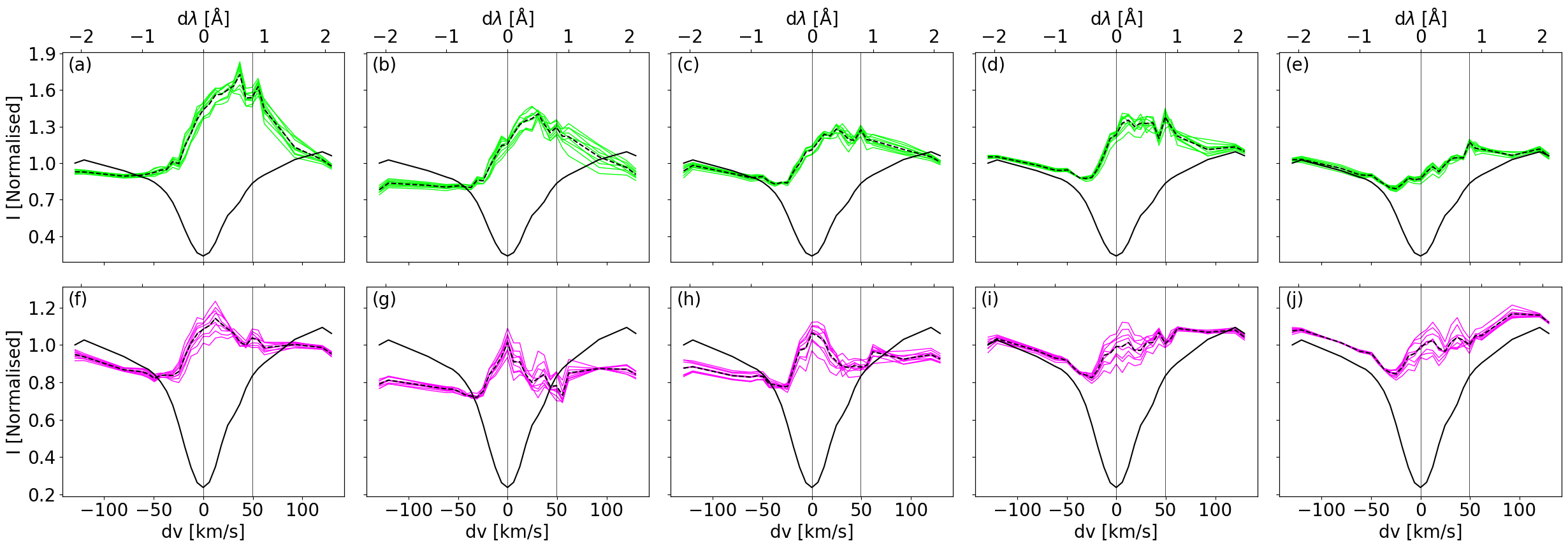}
    \centering
    \caption{
    H$\beta$ profiles obtained from blob and between blob locations.
    The pixel locations are marked in panel~\ref{fig:hbeta_siiv_blob_6panel}f with corresponding colours and labels.
    The left vertical black line shows the line core position and the right vertical line is at $+0.8$~\AA, which has been used for the images in Fig.~\ref{fig:hbeta_siiv_blob_6panel}.
    }
    \label{fig:hbeta_blobs_profiles_from_6panel}
\end{figure*}

We have analysed the spectral line profiles at the locations of the blobs and the regions between the blobs (Fig.~\ref{fig:ha_caii_blobs_colourmap}).
The $3\times3$ grid of profiles in all panels share similar shapes, which indicates that the fine-scale events have been resolved on a \tbold{satisfactory} level.
Intensity variations near the core, particularly in Figs.~\ref{fig:caii_blob_profile}a--b, between adjacent pixels are evident.
The core intensities of the blob profiles are not significantly enhanced compared to other locations along the ribbon.
The blue wings in all panels show little variation between the adjacent pixels and between the different locations, suggesting no upflowing material.
Line broadening is evident in all panels compared to the black solid quiet Sun (QS) averaged profile, noting that broadening effects are not unique to the blob locations.
The blob profiles are differentiated from the other profiles along the ribbon mainly because the former includes an excessive red wing component located at about 20~km s$^{-1}$.
\tbold{Positive d$\nu$ corresponds to red-shifts, and the red wing asymmetry could be a result of downflowing material that produces Doppler-shifted components along the line-of-sight.}
The H$\alpha$ profiles located co-spatially with the \ion{Ca}{II}~8542~\AA\ profiles are shown in Fig.~\ref{fig:ha_blob_profile}.
Similar to the \ion{Ca}{II}~8542~\AA\ profiles, variations in the blue wing tend to be low and consistent between neighbouring pixels.
The general structure of these profiles is more complex as plateaus around the core are evident.
Thus, the plateau-forming profiles envelop the core reversal and make it difficult to determine the position of the core; hence, Doppler shift measurements are impossible.
Regardless of the complexity of the H$\alpha$ profiles, a red wing component is manifested in all the selected profiles.
While the \ion{Ca}{II}~8542~\AA\ profiles have a red component more prominent in the blob pixels, H$\alpha$ consistently exhibits a red component along the flare ribbon, regardless of an evident bright feature in the H$\alpha$ +0.8~\AA\ channel.

\subsubsection{Spectral line profiles in H$\beta$ observations}
\label{sec:chromis_profiles}

We explored the spectral profiles for the blobs observed in H$\beta$ with CHROMIS (see cyan contours in Fig.~\ref{fig:hbeta_siiv_blob_6panel}) in a similar way as that described for \ion{Ca}{II}~8542~\AA\ and H$\alpha$ in Sect.~\ref{sec:crisp_profiles}. 
These profiles are presented in Fig.~\ref{fig:hbeta_blobs_profiles_from_6panel}. 
There is evidence of minor variations near the core and in the red wing for neighbouring pixels, but the consensus is that they do share similar shapes at all locations.
We argue that the observed blobs are resolved at a reasonable level, with the highest spatial and temporal resolution obtained for this observational dataset from SST. 
All profiles along the ribbon are in emission.
Strong red-shifted components are causing the intensity peaks of the blob profiles (green curves) to be located in the red wings.
Determining the peak of each profile is intricate, but a rough estimate suggests the Doppler velocities range between 25--40~km~s$^{-1}$.
This is in contrast to the blobs in \ion{Ca}{II}~8542 +0.5~\AA\ in Fig.~\ref{fig:caii_blob_profile} for which the peaks are near the core rest wavelength, but it is plausible that smaller Doppler shifted components are forming red asymmetric profiles.
Broadening of the blob profiles is evident compared to the average QS black solid line, especially extending toward the red wing.
This broadening is also noticeable in the profiles at the location between the blobs (see panels~(f), (i) and (j)), though not as prominent.
The highest intensity value of the blob profiles had a range estimate from 1.1--1.8 times the average QS blue wing, while profiles at regions between the blobs were confined at 1.0--1.2 times the average QS blue wing.
The profiles in panels~(b)--(d) have approximately similar peak intensities, and their associated blob widths are approximately equal.
A weaker bright point detected as a blob by the FWHM method is labelled as (e) in Fig.~\ref{fig:hbeta_siiv_blob_6panel}b and the profiles are presented in Fig.~\ref{fig:hbeta_blobs_profiles_from_6panel}e.
The core intensities of these profiles are relatively weak compared to the other blob profiles.
However, these profiles consist of a red wing enhancement that is of comparable strength to the other blobs.
This blob was not detected by the algorithm in the \ion{Ca}{II}~8542 +0.5~\AA\ channel but can be identified at the same location at $[X,Y]= [-174\farcs 3, -234\farcs 1]$.
This may be attributed to the higher spatial and temporal resolution of CHROMIS (H$\beta$) than CRISP (\ion{Ca}{II}~$8542$~\AA\ and H$\alpha$).
%



\section{Discussion}
\label{sec:discussion}

We have analysed a GOES C2.4-class flare that was associated with a failed filament eruption triggering two subsequent flares. 
The filament seems to untwist during the eruption as the plasma propagates in a helical motion, as seen in the AIA context movie that is available online. 
During the impulsive phase of the analysed flare, two parallel J-shaped flare ribbons were identified and connected by fibrils.
These J-shaped structures suggest that reconnection \tbold{occurs} in the corona below an erupting flux rope \citep{1996JGR...101.7631D, 2014ApJ...788...60J}.
Additionally, a smaller ribbon was identified south of the eastern ribbon\tbold{, which was} also connected by fibrils.
The fibrils connecting a region of negative photospheric polarity to two regions associated with positive polarities suggest a magnetic field topology where a QSL could be located around the eastern ribbon.
The QSL suggests that the reconnection is associated with a flare current sheet at higher altitudes, presumably at coronal heights.
The data presented by CRISP and CHROMIS reveal fine-scale structures within the ribbons at an unprecedented level, which are comparable with flare kernels observed by IRIS in the \ion{Si}{IV}~1400~\AA\ channel. 
We aimed to identify and trace these chromospheric blobs to evaluate the plasma dynamics.
However, tracking individual features before 08:07:02 is unattainable.
This could be due to the very fast motion of the individual \tbold{features or} that the local region is in the process of being energised before the blobs become prominent.
The blobs likely result from the same reconnection region due to the adjacent and \tbold{co-temporal} formation of the features.
We \tbold{emphasise} the detection of \tbold{the} blobs with a clear periodicity pattern along the flare ribbon, which was supported by multiple frames.
\cite{2021ApJ...920..102W} analysed spatially periodic fine-scale structures in flare ribbons from an analytical 3D magnetic field model.
The study showed that spiral or wavelike structures were formed in the ribbon as a response to the current sheet tearing. 
\tbold{We do not resolve these kinds of fine-structure in our observations, which can be attributed to the resolution limit of the instrument.}
Still, the periodicity \tbold{that we observe} seems to be of a similar nature as \tbold{in} the analytical study \tbold{by \citet{2021ApJ...920..102W}} which suggests our observations \tbold{are} evidence of current sheet tearing.
\tbold{
Although the interpretation of the analytical model provides an explanation for a periodic pattern in the lower atmosphere, it does not describe the energy transport from the current sheet, which is usually explained by electron beams that follow the magnetic field down to the chromosphere.
\cite{2020ApJ...900...18K} modelled the spatial distribution of chromospheric ribbon intensity, assuming energy transport by electron beams and used the resulting radiation to synthesise ribbons from a 3D flare AR.
Radiation from a (single) 1D RADYN simulation \citep{1992ApJ...397L..59C, 2015ApJ...809..104A} was assigned to spatial locations along a number of field lines selected from an extrapolated magnetic field in order to project the predicted emission into a 2D plane and compare the synthetic flare ribbons to real images.
The synthesised ribbons show spatial structure, arising from the timing of the loop energisation -  but also from the set of extrapolated loops chosen and the pixel scale and magnetic field tracing carried out to generate the synthetic images.
Although their simulation had only one set of beam parameters, it is to be expected that a stronger or weaker beam would result in brighter or dimmer ribbon emission, leading to a blob structure.
The reason for a stronger or weaker beam is likely to be related to the coronal energy release, hence, the structures present in the coronal field, the magnetic topology, and the magnetic field evolution are still likely to have an important role in the ribbon fine-structure.
}

\tbold{In our case, multiple} bright points were detected along the ribbon in \ion{Si}{IV}~1400~\AA\ as seen in Fig.~\ref{fig:hbeta_siiv_blob_6panel}c.
These points approach the resolving limit of IRIS but were found to be spatially quasi-periodic during the impulsive phase.
\cite{2021ApJ...920..102W} suggested a linkage between the fine structure in the ribbon with oblique tearing in a current sheet.
Their study was motivated by transient wavelike and spiral substructures seen in the September 2014 X-class flare ribbons in the IRIS \ion{Si}{IV}~1400~\AA\ SJI channel.
Their analytical model gave them control over simulating the tearing by introducing different levels of twist forming small-scale flux ropes.
They identified J-shaped flare ribbons that are associated with a large factor Q, which represents the deformation of the magnetic field mapping in the low solar atmosphere.
%
%
Such deformation of magnetic field mapping, or strong factor Q, could plausibly be identified by fibrils in, for example, the \ion{Ca}{II}~8542~\AA\ channel in Fig.~\ref{fig:sdo_sst_iris_aligned}e.
The eastern ribbon, which is overlying a region of negative polarity, shows fibrils connecting to a region to the west and also to a region to the south where positive magnetic fields are evident.
Additionally, the observed flare ribbons are J-shaped, which, again, is consistent with the magnetic field topology of the model. 
\tbold{Changes in the vertical magnetic field in the photosphere between the pre-flare and post-flare phases were not significant which is in line with previous studies \citep[see, e.g.,][]{2012ApJ...748..138K, 2023ApJ...954..185F},  while this is not as clear for the C-class flare observed by \citet{2021A&A...649A.106Y}.}
%
%
We argue that the underlying blobs in the low chromosphere must be related to patchy reconnection in the coronal current sheet seen as quasi-periodic UV signatures in \ion{Si}{IV}~1400~\AA\ channel and that tearing mode instability is a good candidate for the formation of blobs.
%

%
We have utilised the spectrometry of CRISP and CHROMIS to analyse chromospheric profiles associated with the locations of blobs and between the blobs detected from images.
The \ion{Ca}{II}~8542~\AA\ line is formed at a lower maximum height compared to H$\beta$ \citep{2022A&A...659A.186B}.
While both channels expose blobs, their spectral signatures are different.
A consensus of the blobs from the \ion{Ca}{II}~8542~\AA\ profiles is a secondary component in the red wing at about 20~km~s$^{-1}$ while the cores stay un-shifted.
Profiles at regions between the blobs are also in emission but do not show a secondary component, as seen in Fig.\ref{fig:caii_blob_profile}e-h.
The blob profiles seen in the H$\beta$ +0.8~\AA\ channel are all red-shifted with an estimated Doppler velocity between 25--40~km~s$^{-1}$.
Red-shifted H$\beta$ profiles are accompanied by \tbold{H$\alpha$ profiles with extended red wings that are relatively flat around the core and possibly asymmetric towards the blue.
\cite{2015ApJ...813..125K} modelled electron beams and synthesised H$\alpha$ profiles that show initial upflows at the formation height of the core.
They produced red-asymmetric profiles that were caused by a change in the velocity field below the core-formation region, indicating that blue photons are preferentially absorbed.}
%
\tbold{The authors} concluded that red asymmetric H$\alpha$ profiles are not necessarily associated with down-flowing material during the impulsive phase, also known as chromospheric condensation \citep{1984SoPh...93..105I, 1990ApJ...363..318C}.
This means that the red asymmetric H$\alpha$ profiles alone can not imply chromospheric condensation.
With the support from the red-shifted H$\beta$ profiles, we argue that chromospheric condensation from this event is indeed plausible and is contributing to the formation of red symmetric H$\alpha$ profiles.
Only a single blob may be identified in the H$\alpha$ +0.8~\AA\ channel at a location where also \ion{Ca}{II}~8542~+0.8~\AA\ and H$\alpha$ +0.8~\AA\ show a similar feature, still with temporal differences between the scans.
The representative ribbons in Fig.~9 from \cite{2021ApJ...920..102W} present tight and round spirals based on certain formation heights of the flux ropes in addition to less defined and wavy structures at other formation heights.
The different shapes in different channels may be caused by these effects.
We argue that the blobs are likely formed throughout the chromosphere.
Red wing enhancements from the bright points in H$\beta$~+0.8~\AA\ were also evident in the frame before the detected blobs.
This frame is closer in time to the detected blobs in \ion{Ca}{II}~8542~\AA.
We note that the red wing enhancements in H$\beta$ generate profile peaks in the range 25--40~km~s$^{-1}$ while the red wing component in \ion{Ca}{II}~8542~\AA\ is located approximately at 20~km~s$^{-1}$.
Assuming the red wing enhancements in H$\beta$ are associated with Doppler shifts, that is downflows, a decrease of the vertical component of the velocity at lower altitudes must exist somewhere between the formation heights of H$\beta$ and \ion{Ca}{II}~8542~\AA.
This also puts constraints on plasma compression since velocities are only evident in one line unless an extreme decrease in vertical velocity is present.
Contradictory to this, the red component in the \ion{Ca}{II}~8542 +0.5~\AA\ blob profiles could be formed from a secondary Doppler shifted component.
The peak of the secondary components would be less Doppler shifted than the estimated H$\beta$ peaks, that is $<$25~km~s$^{-1}$, which still indicates a 
decrease in vertical velocity between the formation heights.
\tbold{To our knowledge, the studies on the formation of H$\beta$ during flares from numerical modelling are scarce and, so far, mainly based on 1D simulations.
However, \cite{2017ApJ...850...36C} did synthesise H$\beta$ profiles shown in their Fig.~17 with a central reversal which we did not observe.}
The observational evidence of red-shifted profile peaks \tbold{shows} a change in vertical velocity in the chromosphere between \ion{Ca}{II}~8542~\AA\ and H$\beta$ formation heights.

\tbold{
\cite{2015ApJ...813..125K} analysed red asymmetric \ion{Ca}{II}~8542~\AA\ profiles associated with strong energy deposition from an electron beam.
The red asymmetry was not as prominent as in our observations.
They found bidirectional flows at approximately 1--2~km~s$^{-1}$ in both directions at the formation height of the \ion{Ca}{II}~8542~\AA\ line core.
The flows were associated with extended blue wing and red asymmetry in the line core, which is not in agreement with our observations. 
However, we note that \citet{2015ApJ...813..125K} analyse spectral averages across a small region while our data enables the analysis of individual pixels so that we can better discriminate between blobs with pronounced asymmetries and their surroundings. 
}


\section{Conclusions}
\label{sec:conclusions}

We \tbold{finalise} our work with a summary of the main observational findings, which are as follows.

\begin{enumerate}
    \item We presented high-resolution observations of a C2.4 class flare obtained by SST, IRIS and SDO, which reveal the flaring condition in all atmospheric layers. The AR is composed of a single sunspot with a complex surrounding magnetic field configuration. The chromospheric fibrils related to the flaring atmosphere suggest a QSL topology. The flare was located on the outskirts of the sunspot and formed two prominent ribbons above strong opposite magnetic fields. 
    \item Unique observations by CRISP and CHROMIS revealed fine-scale structures within the ribbon that are referred to as blobs. The measured widths of the blobs vary from 140--200~km and the distances between them vary from 330--550~km in the H$\beta$ images. Although hampered by the variation in seeing conditions, we estimate the lifetime of the blobs to be less than 7~s with an apparent velocity of 11~km~s$^{-1}$. The spatial periodicity of the blobs suggests a response to fragmented reconnection in the coronal current sheet that is likely associated with the tearing mode instability.
    \item At the location of the blobs as seen in the H$\beta$ line, fine-scale brightenings are present in the \ion{Si}{IV}~1400~\AA~SJI channel, too, although the identification of blob-like structures in this channel remains vague due to the lower spatial resolution.
    \item The spectral analysis of the ribbon shows that the blobs along the ribbon show emission lines with a component towards the red wing. In the case of the \ion{Ca}{II}~8542~\AA\ blobs, this component consists of an excessive enhancement at about $+20$~km\,s$^{-1}$. The profiles at regions between the blobs are absent of such a secondary component. The H$\beta$ blob profiles show emission peaks that are red-shifted between 25--40~km~s$^{-1}$.
\end{enumerate}


\begin{acknowledgements}
    We wish to thank the referee for the valuable comments.
    We wish to thank Hugh Hudson for many great discussions in which his knowledge of observational flares has been much appreciated.
    This research is supported by the Research Council of Norway, project number 325491, 
    through its Centres of Excellence scheme, project number 262622, 
    and the European Research Council through the Synergy Grant number 810218 (``The Whole Sun'', ERC-2018-SyG).
    L.F. acknowledge support from grant ST/X000990/1 made by UK Research and Innovation’s Science and Technology Facilities Council.
    G.A. acknowledges financial support from the French national space agency (CNES), as well as from the Programme National Soleil Terre (PNST) of the CNRS/INSU also co-funded by CNES and CEA.
    The Swedish 1-m Solar Telescope is operated on the island of La Palma by the institute for Solar Physics of Stockholm University in the Spanish Observatorio del Roque de los Muchachos of the Institutio de Astrofisica de Canarias.
    The Swedish 1-m Solar Telescope, SST, is co-funded by the Swedish Research Council as a national research infrastructure (registration number 4.3-2021-00169).
    IRIS is a NASA small explorer mission developed and operated by LMSAL with mission operations executed at NASA Ames Research center and major contributions to downlink communications funded by ESA and the Norwegian Space Centre.
\end{acknowledgements}


%
%

\bibliographystyle{aa} 
\bibliography{ads_ref_list} 

\appendix


\end{document}